\documentclass[12pt]{article}
\usepackage{amsmath}
\usepackage{amssymb}
\usepackage{graphicx}
\usepackage{epsfig}
\usepackage{verbatim}
\advance\parskip 1.0pt plus 1.0pt minus 1.0pt   % needs some stretchability
\newcommand{\tr}{\operatorname{tr}}

\newcommand{\be}{\begin{equation}}
\newcommand{\ee}{\end{equation}}
\newcommand{\bea}{\begin{eqnarray}}
\newcommand{\eea}{\end{eqnarray}}
\newcommand{\bA}{\begin{array}}
\newcommand{\eA}{\end{array}}
\newcommand{\bc}{\begin{center}}
\newcommand{\ec}{\end{center}}

\newcommand{\ra}{\rightarrow}

\newcommand{\p}{\partial}
\newcommand{\ie}{{\it i.e.}}
\newcommand{\eg}{{\it e.g.}}

\newcommand{\Nf}{${\cal N}{=}4$}
\newcommand{\Nt}{${\cal N}{=}2$}
\newcommand{\No}{${\cal N}{=}1$}

\def\BC{{\mathbb C}}
\def\BP{{\mathbb P}}

\def\BZ{{\mathbb Z}}

\def\BN{\mbox{\boldmath$N$}}

\begin{document}

\begin{titlepage}
%\vspace{10mm}

\bc

%\hfill  {HUTP-07-xx} \\
\hfill  {TIFR-TH-07-01} \\
\hfill  {\tt hep-th/0701189} \\
         [22mm]

{\Huge Mesonic chiral rings in Calabi-Yau cones \\[6pt] from field theory}
\vspace{10mm}

{\large Lars Grant$^{a, b}$ and K.~Narayan$^a$} \\
\vspace{3mm}
{$^a$\small \it Department of Theoretical Physics, \\
\small \it Tata Institute of Fundamental Research, \\}
{\small \it Homi Bhabha Road, Colaba, Mumbai - 400005, India.\\}
\vspace{1mm}
{$^b$\small \it Jefferson Lab, Department of Physics,\\}
{\small \it Harvard University, \\}
{\small \it Cambridge, MA 02138, USA.\\}
{\small Email: \ lgrant@fas.harvard.edu, \ narayan@theory.tifr.res.in}\\

\ec
\medskip
\vspace{20mm}

\begin{abstract}
We study the half-BPS mesonic chiral ring of the ${\cal N}=1$
superconformal quiver theories arising from $N$ D3-branes stacked at
$Y^{pq}$ and $L^{abc}$ Calabi-Yau conical singularities. We map each
gauge invariant operator represented on the quiver as an irreducible
loop adjoint at some node, to an invariant monomial, modulo
relations, in the gauged linear sigma model describing the
corresponding bulk geometry. This map enables us to write a
partition function at finite $N$ over mesonic half-BPS states. It
agrees with the bulk gravity interpretation of chiral ring states as
cohomologically trivial giant gravitons. The quiver theories for
$L^{aba}$, which have singular base geometries, contain extra
operators not counted by the naive bulk partition function. These
extra operators have a natural interpretation in terms of twisted
states localized at the orbifold-like singularities in the bulk.
\end{abstract}

\end{titlepage}

\newpage
{\small \begin{tableofcontents}
\end{tableofcontents}
}

%\vspace{10mm}

\section{Introduction}

It is of considerable interest to explore strong coupling gauge
theory dynamics, especially in cases of reduced supersymmetry. In
particular, supersymmetric (BPS) states are frequently subject to
non-renormalization theorems and calculations done at weak coupling
in gauge theory can be extrapolated to arbitrary coupling. If the
gauge theory has an AdS/CFT dual, then these calculations can be
matched to corresponding calculations done at strong coupling in the
gravity dual. Various families of nontrivial supersymmetric \No\
gauge theories are obtained by placing D3-branes at nontrivial
supersymmetric conical singularities beginning with \eg\
\cite{dgm97, klebanovwitten, morrisonplesser}, giving rise under
appropriate decoupling limits to families of AdS/CFT dualities,
involving Sasaki-Einstein 5-metrics. More recently, the discovery of
new explicit Sasaki-Einstein 5-metrics \cite{gauntmartspa0403} has
sparked the identification of new families of AdS/CFT dualities
\cite{ypqMartSpark, ypqHanany, Benvenuti:2005ja, Butti:2005sw,
Franco:2005sm}, featuring \No\ superconformal quiver gauge theories
obtained from D3-branes stacked at the tips of families of toric
Calabi-Yau cones. This has been developed further in \eg\
\cite{Bertolini:2004xf, Hanany:2005hq, Hanany:2005ve, Franco:2005rj,
Martelli:2005tp, Benvenuti:2005cz, Butti:2005vn, Butti:2005ps,
Hanany:2005ss, Feng:2005gw}.

The ${1\over 2}$-BPS spectrum of these gauge theories contains
``mesonic'' operators, which have transparent dual interpretations
in terms of D3-brane giant or dual giant gravitons \cite{giantg}
propagating in the dual bulk $AdS_5\times X^5$ spacetime, $X^5$
being the Sasaki-Einstein space in question. In a sense, the
question of understanding these states is similar to that of
understanding the ${1\over 8}$-BPS spectrum of the maximally
supersymmetric \Nf\ SYM theory. In this context, it was shown in
\cite{mikhailov} that D3-branes (with no worldvolume gauge field or
fermion excitations) wrapping surfaces in $S^5$ defined by the
intersections with $S^5$ of the zero sets of arbitrary holomorphic
functions in $\BC^3$ (thought of as a cone over $S^5$) are ${1\over
2}$-BPS classical giant gravitons in $AdS_5\times S^5$. The
quantization of these classical solutions was discussed in
\cite{beasley0207}, and studied more elaborately in
\cite{minwalla0606}, where the gauge theory partition function over
${1\over 8}$-BPS states obtained in \cite{minwalla0510} (see also
\cite{Romelsberger:2005eg}) was recovered.  This provides evidence
that the quantization of BPS solutions gives a sensible subspace of
supersymmetric states of the full Hilbert space. The partition
function of \cite{minwalla0510} can also be recovered by studying
${1\over 8}$-BPS dual giant gravitons \cite{mandal0606}, rendering
further support for duality between giants and dual giants. In the
class of theories we consider here, the Sasaki-Einstein space $X^5$
has a non-contractible 3-cycle, a feature absent in \eg\ the
5-sphere $S^5$. This gives rise to ``baryonic'' operators in the
gauge theory, dual to giant gravitons corresponding to D3-branes
wrapped on the 3-cycle. The quantization of these baryons was
studied in \cite{beasley0207} in the context of the conifold.

The study of the BPS spectrum in the context of the \No\ theories
obtained from D3-branes on Calabi-Yau cones has been developed further
by the plethystic program of \cite{Benvenuti:2006qr, Feng:2007ur},
(see also \cite{Hanany:2006uc} for \Nt\ theories) as well as
\cite{Butti:2006au} (who in particular study baryons). The mesonic BPS
spectrum has also been studied using dual giants in
\cite{Martelli:2006vh, Basu:2006id}.

In this paper, we present a slightly different field theory approach
to studying the BPS spectrum in these \No\ theories, preliminary
results of which were reported in \cite{Strings06}. This is based on
the fact that the geometries in question are toric, admitting
descriptions in terms of gauged linear sigma models (GLSMs)
\cite{wittenPhases} (developed further for toric varieties by
\cite{morrisonplesserInstantons}). We map
the chiral ring of gauge invariant mesonic operators in the gauge
theory represented as irreducible (closed) loops based at any one
node on the quiver to a basis set of invariant monomials (modulo
relations) in the corresponding GLSM (Sec.~\ref{sec:Ypq}). We find a
minimal generating basis for the chiral ring using the F-term
equations of motion in the quiver to eliminate redundancies. This
field theory analysis is closely tied to the holomorphic quotient
construction of the geometry, which describes the toric variety as a
set of invariant monomials satisfying relations. Since the basis set
of invariant monomials generates the ring of holomorphic functions
on the toric variety, this enables us to write out the partition
function over ${1\over 2}$-BPS mesonic states in the gauge theory,
thus agreeing with their interpretation in terms of giant gravitons
propagating in the bulk $AdS_5\times L^{abc}_5$ spacetime.  From the
bulk point of view, generalizing \cite{mikhailov}, the set of
cohomologically trivial giant gravitons in these theories
(restricting first to smooth base geometries) is simply given by the
intersection with the base of arbitrary holomorphic functions on the
toric variety defined by the Calabi-Yau cone over the base: these
can be described using the GLSM, which provides a natural
(equivalent) symplectic quotient description of the geometry.  Our
procedure suggests a correspondence between the underlying
coordinates of the geometry/GLSM and bifundamentals in the quiver,
which is useful in studying baryonic operators using open paths on
the quiver: these baryons are D3-branes wrapped non-trivially on the
noncontractible 3-cycle in the base. We hope to report further
progress on this later \cite{wip}. We also study $L^{abc}$s with
singular base geometries (Sec.~\ref{sec:sing}). In this case, we
find that there are extra operators in the quiver theory which
cannot be generated by the analogs of the irreducible loop
generators in the quiver. These extra operators correspond to the
singular loci in the base geometries, suggesting a bulk
interpretation for these operators in terms of localized or twisted
giant gravitons, analogous to twisted closed string states localized
at orbifold singularities. The bulk partition function above
specialized to these singular spaces does not capture these
localized giant gravitons: it would be interesting to develop a
deeper understanding of these states, with possible generalizations
of \cite{mikhailov}.

\section{The $L^{abc}$ geometries}

Several nonspherical horizon generalizations of the AdS/CFT
correspondence appeared in \cite{morrisonplesser}, whose authors
constructed gauge theory duals for D-branes at various nontrivial
singularities by starting with a known orbifold quiver
\cite{douglasmoore, dgm97} theory and turning on appropriate
Fayet-Iliopoulos parameters to flow down to the gauge theories in
question. In principle such techniques may be used to ``derive'' the
\No\ quiver gauge theory duals on the classes of toric Calabi-Yau
conical singularities we discuss here. These involve bulk $AdS_5\times
L^{abc}_5$ spacetimes with $N$ units of five form flux, the
$L^{abc}_5$ being the Sasaki-Einstein base manifolds for the
$L^{a,b,c}$ singularities, specified by three positive integers
$(a,b,c)$, as we discuss below.  The quiver gauge theory duals arising
on D-branes located at the singular tips of the cones can be described
explicitly, as in \cite{Franco:2005sm} who construct them using brane
tiling and toric geometry techniques.

Now we describe the $L^{abc}$ geometries in greater detail. Consider a
2-dimensional $(2,2)$ worldsheet supersymmetric gauged linear sigma
model (GLSM) \cite{wittenPhases, morrisonplesserInstantons} (with zero
Fayet-Iliopoulos parameter) with four chiral superfields $z_i$ and a
single $U(1)$ gauge field with gauge transformations given by a charge
matrix $Q$
\bea \label{labcQ}
Q&=&(\bA{cccc} a & b & -c & -d \eA)\ ,\ \ \ a,b,c,d > 0\ ,\ \ \ \sum_iQ_i=0\ ,
\nonumber\\
z_i &\ra& e^{i Q_i \alpha} z_i\ , \qquad {\rm with\ real}\ \alpha\ .
\eea
The low energy dynamics of this 2-dimensional field theory is a sigma
model on a (noncompact) Calabi Yau space that may be thought of as the
submanifold
\be \label{dterm}
\left\{ \sum_i Q_i |z_i|^2=0 \right\} // U(1)\ ,
\ee
which, in the GLSM, is the D-term equation modulo the $U(1)$ gauge
equivalence. This is the symplectic quotient description of these spaces.

We can also describe these spaces via a holomorphic quotient
construction, in general describing the space as a collection of
monomial relations\footnote{Some aspects of the geometry are
discussed in \cite{knconiflips}, which studies the dynamics of
unstable nonsupersymmetric conifold-like singularities described by
GLSMs with charge matrix $Q = (\bA{cccc} n_1 & n_2 & -n_3 & -n_4 \eA)$.
The $L^{abc}$s and $Y^{pq}$s are the supersymmetric subclass
($\sum_iQ_i=0$) in these.}.
Let $F$ represent the union of the surfaces $z_a=z_b=0$ and $z_c=z_d=0$.
Let $\BC^*$ represent the complexified $U(1)$ gauge action
$z_i\ra\lambda^{Q_i} z_i$ where $\lambda\in\BC^*$ is an arbitrary complex
number. Then the Calabi-Yau space is described as the toric variety
$\frac{\BC^4-F }{\BC^*}$, where $F$ is the excluded set. The equivalence
of this description with that of the previous paragraph may be
demonstrated by fixing the real part of the $\BC^*$ gauge invariance
($z_i\ra \lambda^{Q_i} z_i$ for real $\lambda$) and solving \eqref{dterm}.
The simplest such singularity is of course the supersymmetric
conifold, $Q = ( \bA{cccc} 1 & 1 & -1 & -1 \eA )$, with a basis of
gauge invariant monomials given by\
$x_1=z_az_c, \ x_2=z_az_d, \ x_3=z_bz_c, \ x_4=z_bz_d$, as we have
seen earlier. These satisfy the relation
%\be\label{1111}
$x_1x_4=x_2x_3$\ ,
%\ee
which describes the conifold as a 3-complex dimensional hypersurface
in $\BC^4[x_i]$. In general, the $L^{abc}$ spaces are not complete
intersections of hypersurfaces, \ie\ the number of variables minus
the number of equations is not equal to the (complex) dimension,
\ie\ 3, of the space. For example, the singularity\ $Q = ( \bA{cccc}
1 & 3 & -2 & -2 \eA )$, which is the space $Y^{21}$, has a monomial
basis\ $x_1=z_a^2z_c, \ x_2=z_a^2z_d, \ x_3=z_b^2z_c^3, \
x_4=z_b^2z_d^3,\ x_5=z_az_bz_cz_d, x_6=z_az_bz_c^2,\
x_7=z_az_bz_d^2,\ x_8=z_b^2z_c^2z_d,\ x_9=z_b^2z_cz_d^2$. One can
check that there are at least 9 relations here\ $x_1^3x_4=x_2^3x_3,\
x_1x_4=x_5x_7,\ x_2x_3=x_5x_6,\ x_1x_3=x_6^2,\ x_2x_4=x_7^2,\
x_8x_9=x_3x_4,\ x_6x_7=x_1x_9=x_2x_8$.

We now return to the symplectic quotient description. Consider the
intersection of our Calabi Yau space with the 7-sphere
$\sum_iQ_i|z_i|^2=r^2$ in $\BC^4$. At $r=1$, the 5 real dimensional
space so obtained is, by definition, the Sasaki-Einstein space
$L^{abc}$. The metric on the Calabi Yau space takes the form\ \
$ds^2=dr^2+ r^2 ds_{L^{abc}_5}^2$.\ In other words the Calabi Yau
space is a cone whose base is the Sasaki-Einstein manifold
$L^{abc}_5$. As $L^{abc}_5\neq S^5$, the tip of this cone is
singular. From the GLSM point of view, the full space described as
the symplectic quotient at the singular point\ \
$\left\{a|z_a|^2+b|z_b|^2=c|z_c|^2+d|z_d|^2\right\} //U(1)$, \ can
be understood as a cone over a 5D base by looking at a cross section
of the cone at some finite distance from the singularity
($z_i=0,\forall i$), \ie\ \ $a|z_a|^2+b|z_b|^2+c|z_c|^2+d|z_d|^2=1$,
which is a squashed or ellipsoidal $S^7$. This then gives \be
\left\{a|z_a|^2+b|z_b|^2=c|z_c|^2+d|z_d|^2={1\over 2}\right\}
//U(1)\ , \ee so that the base is $\{{\tilde S^3}\times {\tilde
S^3}\}/U(1)$, with ${\tilde S^3}$ denoting an ellipsoidal $S^3$, the
$U(1)$ acting with the charge matrix $Q_i$ given in (\ref{labcQ}).
The $U(1)$ quotienting then gives for the base space, fibrations
involving Lens spaces $S^3/\Gamma$ with appropriate discrete groups
$\Gamma$. For instance, the coordinate patch where $z_a\neq 0$ is
gauge fixed to be real admits a residual gauge symmetry $\BZ_a$
acting on the space giving rise to discrete identifications on
$z_b,z_c,z_d$. Then the $z_b$ part of the space gives a $\BP^1$
alongwith with a Lens space $S^3/\BZ_a$ with identifications\
$(z_c,z_d)\ra (e^{2\pi c/a}z_c,e^{2\pi d/a}z_d)$. There are similar
descriptions on the other coordinate patches $z_i\neq 0$. See also
Sec.~\ref{sec:sing} for various aspects of the geometry, in
particular involving singularities on the base.

\subsection{Dual quivers for the $L^{abc}$ geometries}

\begin{figure}
\bc
\epsfig{file=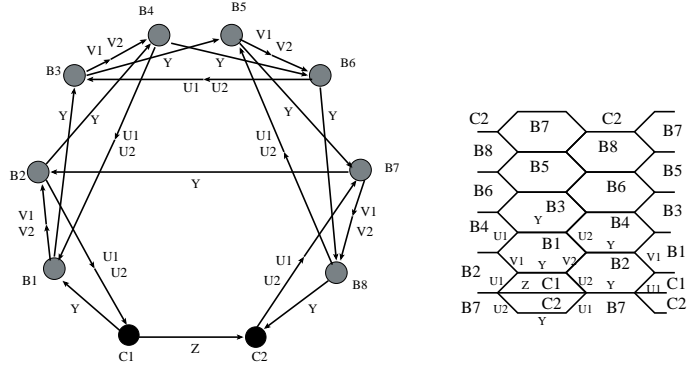, width=9cm}
\caption{The quiver and brane tiling for $L^{195} (=Y^{54})$}
\label{Y54}
\ec
\end{figure}
We now describe some essential features of the $L^{abc}$ quiver theories,
mainly reviewing their construction in terms of brane tilings
\cite{Franco:2005sm}.
The gauge dual to the geometry $AdS_5\times L^{abc}$ is a quiver gauge
theory in which all gauge groups are $SU(N)$.  The global symmetry of
the theory is $U(1)_{F_1}\times U(1)_{F_2}\times U(1)_B\times U(1)_R$.
The $U(1)_{F_i}$ are flavour symmetries and the $U(1)_B$ and $U(1)_R$
are baryon and R-symmetries respectively.
The number of fields in the gauge theory is given by:
\begin{equation}
N_g=a+b \qquad N_f=a+3b
\end{equation}
where $N_g$ is the number of gauge groups \ie\ the number of nodes in the
quiver diagram, and $N_f$ is the number of bifundamental chiral fields,
\ie\ the number of lines in the quiver. There are four classes of chiral
bifundamentals. We call these classes $Y,Z,U_1,U_2$. There are two
additional classes of fields $V_1, V_2$. Fields in a class share the
same Baryon, Flavour and R-charges. The multiplicity of fields in each
class is
\be
mult[Y]=b \quad mult[Z]=a \quad mult[U_1]=d \quad mult[U_2]=c \quad
mult[V_1]=b-c \quad mult[V_2]=c-a\ ,
\ee
\ie\ there are $b$ distinct fields of type $Y$ carrying the same
charges and so on.

There are $2b$ terms in the superpotential. A table listing all
charges and multiplicities of each kind of field is found in
\cite{Franco:2005sm}.
There are 4 distinct types of nodes in the quiver:
\begin{equation}
A=U_1Y V_1\cdot U_1 Y V_1\qquad B=V_2YV_1\cdot U_1YU_2\qquad
C=YZ\cdot U_1U_2\qquad D=U_2YV_2\cdot U_2YV_2
\end{equation}
This notation means that at a $B$ node for example, there are three
bifundamentals, of type $V_2,Y,V_1$ coming in and three bifundamentals
of type $U_1,Y,U_2$ going out or vice-versa.

Calling the multiplicity of each type of node in the diagram
$(n_A,n_B,n_C,n_D)$, we have
\be
n_C=2a\ , \qquad n_B+2n_A=2(b-c)\ , \qquad n_B+2n_D=2(c-a)\ ,
\ee
as relations among these positive integers.  These relations do not
completely fix the quiver. For most values of $(a,b,c)$, there is a
line of solutions to these equations, with different gauge theories
lying on this line related by Seiberg duality.

We can think instead of four tiles associated with these nodes and
draw a brane tiling for the theory. The number of type $C$ tiles will
be $n_C/2$. The other types occur $n_A,n_B$ and $n_D$ times.  Since
there are only 4 kinds of tiles, there are only three kinds of terms
\be
YU_1ZU_2 , \qquad YU_1V_1 , \qquad YU_2V_2\ ,
\ee
in the superpotential, the sign of each term given by the associated
brane tiling \cite{Franco:2005rj, Franco:2005sm}.

To extract the superpotential from the tiling, we colour the
vertices in the brane tiling in alternating black or white. The
terms in the superpotential are given by drawing clockwise loops
around white nodes and anticlockwise loops around black nodes,
contracting the fields on the edges crossed by the loop in the order
determined by the colour of the node and assigning a minus sign to
say black nodes and a positive sign to white nodes.
See \eg\ Figure~\ref{Y54} for a picture of the brane tiling for the
space $L^{195} (=Y^{54})$ and the associated quiver.

\section{$Y^{pq}$ quiver theories}\label{sec:Ypq}

\begin{figure}
\bc \epsfig{file=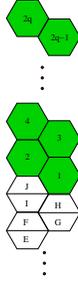, width=1cm}
\caption{The tiling for a
general $Y^{pq}$ space} \label{y94tile}
\ec
\end{figure}
\begin{figure}
\bc \epsfig{file=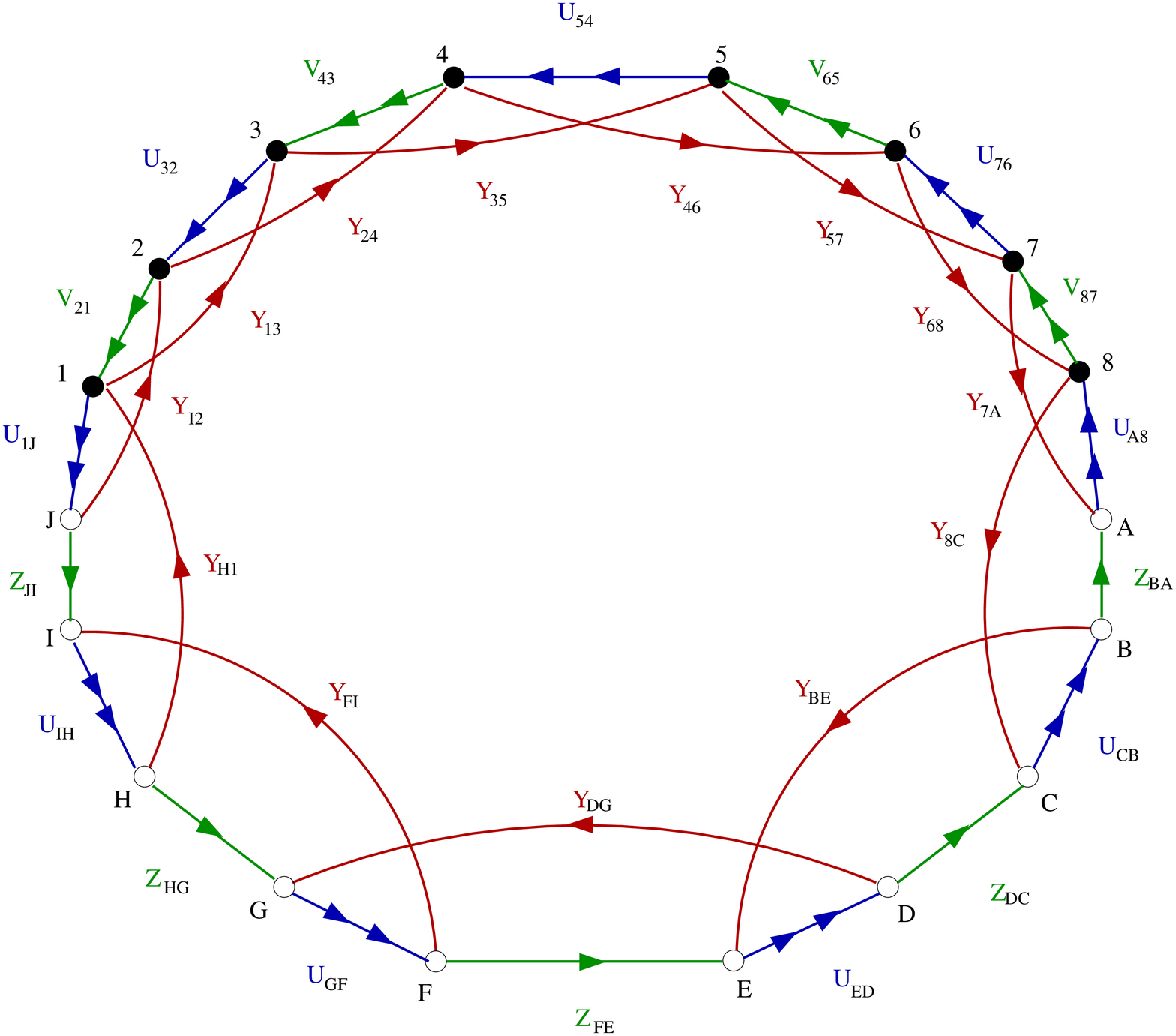, width=7cm} \caption{The quiver for
$Y^{94}$} \label{y94} \ec
\end{figure}

The Calabi-Yau cones $Y^{pq}$ with $p,q$ coprime with $p>q$ are a
smooth subclass of the $L^{abc}$ geometries \cite{ypqMartSpark}. They
are given by taking:
\be
Q=(a,b,-c,-d)=(p-q,p+q,-p,-p)\ , \qquad p,q>0 \ {\rm coprime}\ ,
\ee
$Y^{1,0}$ being the conifold. The conditions on the number of nodes in
the quiver for these $Y^{pq}$ geometries are
\be
n_C=2(p-q)\ , \qquad n_B+2n_A=2q\ , \qquad n_B+2n_D=2q\ .
\ee
Thus we can choose $n_A=n_D=0$ so that we have only $C$ nodes and $B$
nodes, giving $n_C=2(p-q)$ and $n_B=2q$. The tiling for the gauge theory
corresponding to $Y^{pq}$ is shown in Figure~\ref{y94tile}. These
theories have wheel-like quiver diagrams (see Figure~\ref{y94} for the
quiver for $Y^{94}$). The superpotential $W$ in this case has $2(p+q)$
terms which can be read off the nodes of the tiling. The links in the
quiver with one arrow represent single bifundamental fields, while the
links with $k$ arrows represent $k$ fields that are bifundamentals under
the same two gauge groups. The multiplicities of the quiver fields are
\be mult[Y]=p+q\ , \quad
mult[Z]=p-q\ , \quad mult[U_1]=mult[U_2]=p\ , \quad
mult[V_1]=mult[V_2]=q \ .
\ee

In what follows, we discuss mesonic branches of the classical chiral
ring of these $Y^{pq}$ quiver theories. In principle, our methods
here should apply equally well to smooth $L^{abc}$
geometries\footnote{An $L^{abc}$ geometry has a smooth base if each
of $a,b$ is coprime with each of $c,d$: Sec.~\ref{sec:sing}
discusses this smoothness criterion of the $L^{abc}$s in greater
detail.} that may be distinct from the $Y^{pq}$s which are a smooth
subclass.

\subsection{Mesonic sector of the chiral ring}

In the gauge theory, the structure of the moduli space (defined by the
equations of motion, $\partial W=0$) allows us to map a basis of gauge
invariant commuting adjoint operators represented as irreducible loops
(based at any one node) on the quiver to a basis set of $2p+5$
invariant monomials in the GLSM. Then the chiral ring\footnote{See \eg\
\cite{cdsw0211} for a general discussion of chiral rings in \No\
theories.} of gauge
invariant mesonic operators on the moduli space precisely reproduces
the geometry of the singularity described as a toric variety in terms
of invariant monomials satisfying relations. This is a natural
generalization of the realization of the geometry of \eg\ orbifold
singularities from the moduli spaces of D-brane quiver theories
\cite{douglasmoore, dgm97} \footnote{In the course of writing this
paper, we came across \cite{berenstein0505, pinansky}, who use similar
techniques in the context of the del Pezzo surfaces $dP^{1,2}$.}. The
structure of this map has built into it the following correspondence
between the underlying variables of the geometry/GLSM and
bifundamentals in the quiver
\bea\label{mapGLSMquiver}
Y \ \sim\ z_a\equiv a\ , \qquad\qquad Z \ \sim \ z_b\equiv b\ , &&
\qquad\ \ V_1 \ \sim \ z_bz_d\equiv bd\ , \nonumber\\
U_1 \ \sim \ z_c\equiv c\ , \qquad\qquad U_2 \ \sim \ z_d\equiv d\ ,
&& \qquad\ \ V_2 \ \sim \ z_bz_c\equiv bc\ .
\eea
We use this to identify quiver loops and GLSM monomials (we
have for notational convenience, used $a,b,c,d$, for the GLSM fields
and will continue to do so in what follows).
As it stands, this is a many-to-one map, consistent with the baryon
charges in the gauge theory mapping to the $U(1)$ charges in the GLSM.

\begin{figure}
\bc
\epsfig{file=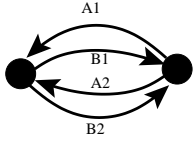, width=2.5cm}
\caption{The quiver for the supersymmetric conifold}
\label{L111}
\ec
\end{figure}
We will first quickly review the familiar conifold \cite{klebanovwitten,
morrisonplesser} from this point of view. The quiver in this case is
shown in Figure~\ref{L111}: the theory has bifundamentals $A_i$
in the $(N,\bar{N})$ and $B_j$ in the  $(\bar{N},N)$ of the
$SU(N)\times SU(N)$ gauge group. The superpotential is\
$W={\lambda\over 2}\epsilon^{ij}\epsilon^{kl}\tr A_iB_kA_jB_l$. The
four equations of motion $\partial_i W=0$ are of the form\
$A_1B_1A_2=A_2B_1A_1$. We define four gauge invariant operators
$x_k\equiv A_iB_j$, which can be identified with the four irreducible
closed loops on the quiver starting at the right node. Then we see
using the equations of motion that these commute and satisfy the
relation\ $x_1x_4=x_2x_3$, which is the familiar equation of the
conifold as a hypersurface in $\BC^4[x_i]$. In what follows, we will
use essentially similar methods in the $Y^{pq}$ theories, where the
incomplete intersection nature of the geometry makes the story
somewhat more complicated.

We begin by discussing the GLSM and associated toric description for
these geometries, first constructing a list of the invariant monomials
modulo relations characterizing these geometries as toric varieties.
There are $2p+5$ invariant monomials in the GLSM
\be
abcd\ , \quad abc^2\ , \quad abd^2\ , \quad a^pc^{(p-q)-k}d^k ,
\quad b^pc^{(p+q)-l}d^l , \qquad k=0\ldots p-q\ ,\  l=0\ldots p+q\ ,
\ee
with some of the relations between these monomials being
\bea \label{apbpreln}
(abc^2)(abd^2)&=&(abcd)^2 \nonumber\\
%\label{abreln}
(a^pc^{(p-q)-k_1}d^{k_1})\
(b^pc^{(p+q)-k_2}d^{k_2})&=&(ab)^pc^{2p-n}d^n=
(abc^2)^{l_1}(abd^2)^{l_2}(abcd)^{l_3}\\
(a^pc^{(p-q)-k}d^{k})(abc^2)&=&(a^pc^{(p-q)+2-k}d^{k-2})(abd^2)
=(a^pc^{(p-q)+1-k}d^{k-1})(abcd)\nonumber \\
(b^pc^{(p+q)-k}d^{k})(abc^2)&=&(b^pc^{(p+q)+2-k}d^{k-2})(abd^2)
=(b^pc^{(p+q)+1-k}d^{k-1})(abcd)\nonumber
\eea
where $k_1+k_2=l_1+l_2+l_3=n$ and $l_3=0$ or $1$.

Now let us move onto the quiver theory. Using the map
(\ref{mapGLSMquiver}) between the quiver gauge fields and GLSM fields,
we can easily identify the invariant loops on the quiver. We label the
nodes on the general $Y^{pq}$ quiver as $B1,\ldots,B2q,C1,\ldots,C2(p-q)$
going around in the clockwise direction. The generators are identified
as follows:
\bea\label{ypqgen}
abcd, abc^2, abd^2 \ & \sim & \
Y_{C2(p-q)B2}V^{1,2}_{B2B1}U^{1,2}_{B1C2(p-q)} \\
a^pc^{(p-q)-k}d^{k} \ & \sim & \ Y_{C2(p-q)B2}Y_{B2B4}Y_{B4B6}\ldots
Y_{B(2q-2)B(2q)}Y_{B2qC3}U^{1,2}_{C3C2}Y_{C2C5}U^{1,2}_{C5C4}Y_{C4C7}
\ldots \nonumber \\
&&Y_{C2(p-q-1)B1}U^{1,2}_{B1C2(p-q)}\nonumber \\
b^pc^{(p+q)-k}d^k \ & \sim & \
Z_{C2(p-q)C2(p-q)-1}U^{1,2}_{C2(p-q)-1C2(p-q)-2}Z_{C2(p-q)-2C2(p-q)-3}
\ldots\nonumber
\\ &&Z_{C2C1}U^{1,2}_{C1B2q}V^{1,2}_{B2qB2q-1}U^{1,2}_{B2q-1B2q-2}
V^{1,2}_{B2q-2B2q-3}\ldots
V^{1,2}_{B2B1}U^{1,2}_{B1C2(p-q)}\nonumber
\eea
These operators are in a sense the obvious generators: a set of
triangles at the node $C2(p-q)$, a set of loops going clockwise and
another going anticlockwise around the wheel.

Now we explain how these operators generate all possible paths on
the quiver diagram. The $F$ equations generated by differentiating
the superpotential with respect to $U$s and $V$s along the outside
of the wheel set neighboring $UVY$ triangles or $UZVY$ squares equal
to each other. But drawing paths on the wheel, we can go around the
wheel in the clockwise direction or anticlockwise direction or
insert triangles and squares. But the $F$ equations allow us to move
all the squares and triangles around to the triangles on the
$C2(p-q)$ node and we are left with a path written as a product of
the generators above. For a more rigorous analysis, see
section~\ref{sec:Y32} where the case $Y^{32}$ is worked out in
detail. The $F$ equations obtained by differentiating with respect
to the $Y$ fields restrict the number of generators so that the
generators in the quiver theory are in one-to-one correspondence
with the invariants in the GLSM.

For convenience in exploiting the equations of motion, we label these
by a concise notation as follows. First we suppress explicitly
labeling the fields including their multiplicities (as in
Figure~\ref{Y54}): for example, both $Y_{C1,B1}$ and $Y_{B1,B3}$ are
labeled $Y$ with the understanding that the specific field will be
clear from the context. Then the set of generating loops (operators)
is
\bea
&& Y11\ , \ Y12\ , \ Y21\ , \ Y22\ , \quad
Y^{q+1}1Y1Y\ldots Y1\ , \ Y^{q+1}1Y2Y\ldots Y2\ , \ etc \nonumber\\
&& {} Z1Z\ldots 1Z \ 11\ 11\ 11\ 11\ 1 \ , \quad Z1Z\ldots 2Z\ 21\ 11\
11\ 11\ 1\ , \ etc\ ,
\eea
where the last set of gauge invariant loops is a potentially large list
of operators. However there are several equations of motion for the
various fields.

The action of these $F$ term constraints on the generators is simple
in this notation. For instance from any of the $Y$s excepting $Y_{B7
B2}$, we have equations of the form\ $V_1U_1=V_2U_2$.  Thus using
the above notation, for two neighbouring fields we have $11=22$.
Similarly the other equations of motion imply that for two fields
separated by a $Y$ or a $Z$ we have $1Z2=2Z1$ and $1Y2=2Y1$. These
F-term equations cut down the number of distinct loops (operators)
so that an arbitrary loop (operator) can be generated by a basis of
loops, which is precisely the set of invariant monomials in the
GLSM. Overall, it can be shown that there are three linearly
independent triangle generators, $p-q+1$ clockwise loops
$Y^{q+1}\ldots$ and $p+q+1$ anticlockwise loops $Z\ldots$. Thus the
independent generators in the quiver theory are in one-to-one
correspondence with the invariants in the GLSM.

The generators in the quiver also commute with each other.  We show
the commutation of the triangles, denoting $Y_{13}=\bar{Y}$ and
$U^{1,2}_{32}$ as $\bar{1}$ or $\bar{2}$.
\begin{eqnarray}
Y12Y21&=&Y1\bar{Y}\bar{2}21=Y1\bar{Y}\bar{1}11=(Y11)^2\\
Y21Y12&=&Y2\bar{Y}\bar{1}12=Y2\bar{Y}\bar{2}22=(Y22)^2=(Y11)^2
\end{eqnarray}
So we have $[Y12,Y21]=0$. The other triangle operators commute in a
similar manner.

Suppose we multiply $triangle \ \times \ loop$, where by loop we
mean one of the clockwise or anticlockwise loops generators listed
above. The $F$ equations allow us to move the triangle around from
the beginning of the loop to the end, so that we have $[\ triangle \
,\ loop\ ]=0$.

Finally if we multiply $clockwise \ \times \ anticlockwise $, we
will have a triangle where the two paths connect. As we move this
triangle anticlockwise around the wheel, we collect more and more
triangles, so that we can write this operator as a product of
triangles: $clockwise \ \times \ anticlockwise \ = \ triangle^p$,
where by $triangle^p$ we schematically represent the product of
various triangle operators. We can also write $anticlockwise \
\times \ clockwise = triangle^p$ and we can show that we can take
the same triangles to occur in both expressions.  Therefore the loop
generators also commute.

The generators in the quiver theory also satisfy the same relations
as the invariants of the GLSM.  We have shown above the relation
satisfied by the triangles.  We have also explained how the relation
$clockwise \ \times  \ anticlockwise \ = \ triangle^p$ comes about.
These relations correspond to the relations (\ref{apbpreln}) in the
GLSM. The relations $loop \ \times \ triangle = loop' \ \times \
triangle'$ also follow from the $F$ term equations.

We see that the quiver gauge theory contains a set of commuting
gauge invariant operators, represented as irreducible loops at a
single node, that are in one-to-one correspondence with the $r
(=2p+5)$ invariant monomials in the GLSM, and satisfy the same
relations as the GLSM invariants (section~\ref{sec:Y32} describes
this explicitly for the $Y^{32}$ case). For an $SU(N)^{N_g}$ theory,
these $r$ commuting operators are complex matrices that can be
upper-triangulated by a gauge transformation\footnote{At generic
points in moduli space, where all eigenvalues of all $r$ matrices are
distinct, we can simultaneously diagonalize these matrices. However
the matrices are complex and may not be diagonalizable if some
eigenvalues are repeated.}. All gauge invariant functions of an
upper-triangular matrix are symmetrized functions of the eigenvalues
alone. Thus the quiver moduli space reproduces the geometry of the
Calabi-Yau cone (or more precisely the symmetric product of $N$
copies of the Calabi-Yau cone, for $N$ branes),\footnote{As an aside, we
note that this can be regarded as a consistency check of the brane
tiling techniques used \cite{Franco:2005sm} to construct the $L^{abc}$
quiver theories.} enabling us to
recover the mesonic chiral ring of the gauge theory. With a view to
doing this, consider first a $U(1)^{N_g}$ theory. The partition
function over mesonic BPS states, using the state-operator map,
counts ``words'' generated by the $2p+5$ invariant monomials $w_j$
in the GLSM \be\label{pf} Z_0=\sum_{m_j} \left(\prod_{j=1}^{2p+5}
w_j^{m_j}\right) \ \equiv\ \sum_{\sum_in_iQ_i=0} \left(\prod_{i=1}^4
z_i^{n_i}\right)\ . \ee The last expression here, involving the
$z_i$, written in terms of the GLSM variables defining the toric
variety, is equivalent to that in terms of the $w_j$ given by the
holomorphic quotient. Replacing the $z_i$ by bosonic creation
operators $a_i^{\dag}$, we see that this is the partition function
for a 4D (bosonic) harmonic oscillator with the constraint
$\sum_in_iQ_i=0$. Since gauge invariant operators of the
$SU(N)^{N_g}$ theory are symmetrized functions of the eigenvalues,
the Hilbert space for this theory is a symmetric product of $N$
copies of that of the $U(1)^{N_g}$ theory. The partition function is
then that of $N$ bosons in a 4D harmonic oscillator potential with
the constraint $\sum_in_iQ_i=0$, and is given by the coefficient of
$p^N$ in \be\label{multipf} Z=\prod_{\sum_in_iQ_i=0}\  {1\over 1-p
e^{-\sum_i n_i \beta_i}}\ . \ee This partition function has appeared
in \cite{Benvenuti:2006qr}.

We now make a few comments on the bulk point of view. The $z_i$ carry
the charges of the fields $Y,Z,U_1,U_2$ respectively.  Thus the chiral
ring of a $Y^{pq}$ quiver theory can be interpreted as the set of all
holomorphic functions in $\BC^4$ which are invariant under the $U(1)$
action $z_i\to e^{iQ_i\alpha}z_i$. For any such function $f(z_i)$,
setting $f(z_i)=0$ gives a holomorphic divisor of the Calabi-Yau
cone. The corresponding giant graviton is then given by the
intersection of this surface with the base, generalizing
\cite{mikhailov}. We have thus recovered in (\ref{multipf}) the bulk
partition function over all such cohomologically trivial, \ie\
mesonic, giant gravitons propagating in the bulk $AdS_5\times
L^{abc}_5$ spacetime. The fact that the gauge theory partition
function agrees with that obtained from the bulk suggests a
non-renormalization theorem for these BPS states.

In what follows we describe in detail the example $Y^{32}$. Similar
analyses can be performed for other $L^{abc}$ spaces with smooth
base 5-geometries, as well as other spaces obtained from \eg\
$\BC^3/\BZ_N$ singularities (which are also toric), and perhaps more
generally.

\subsection{An example: $Y^{32}$}\label{sec:Y32}

\begin{figure}
\bc \epsfig{file=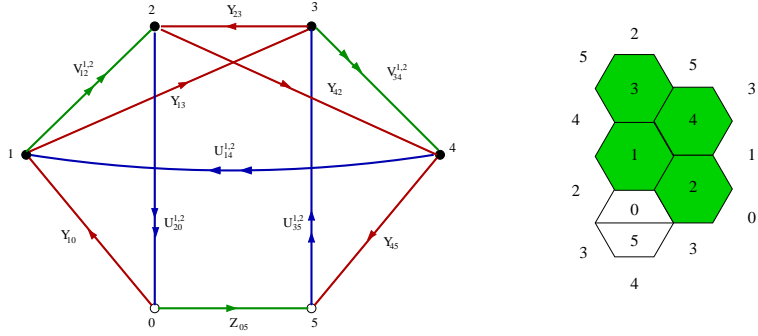, width=10cm} \caption{The quiver and brane
tiling for $Y^{32}$} \label{Y32} \ec
\end{figure}

In this section, we illustrate explicitly the $Y^{32}$ theory,
defined by\ $Q=(1,5,-3,-3)$.  The $2p+5=11$ invariants here are
\begin{eqnarray}
&abcd \qquad abc^2 \qquad abd^2 \nonumber \\
&a^3c \qquad a^3d \\
&b^3c^5 \qquad b^3 c^4d \qquad b^3 c^3 d^2 \qquad b^3 c^2 d^3 \qquad
b^3 c d^4 \qquad b^3 d^5 \nonumber
\end{eqnarray}
The resulting quiver and brane tiling are shown in Figure~\ref{Y32}.
The superpotential is given by:
\begin{eqnarray}
W&=&Y_{32}U_{20}^1 Z_{05}U_{53}^2-Y_{32}U_{20}^2Z_{05}U_{53}^1\nonumber\\
&&\ +Y_{01}V_{12}^2U_{20}^2-Y_{01}V_{12}^1U_{20}^1
+Y_{24}U_{41}^1V_{12}^1-Y_{24}U_{41}^2V_{12}^2\\
&&\ +Y_{13}V_{34}^2U_{41}^2-Y_{13}V_{34}^1U_{41}^1
+Y_{45}U_{53}^1V_{34}^1-Y_{45}U_{53}^2V_{34}^2\nonumber
\end{eqnarray}
Each term here is a triangular or square loop on the quiver diagram.
Every edge in the quiver appears in two such loops which appear with
opposite sign in the superpotential. This makes the constraints
${\partial W\over \partial X}=0$, for any field $X$ in the quiver,
easy to identify on the quiver diagram.

\begin{figure}
\bc \epsfig{file=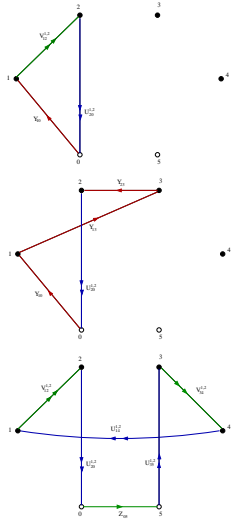, width=3cm} \caption{The basis of
irreducible loops in the $Y^{32}$ quiver, based at node $0$.}
\label{Y32loops} \ec
\end{figure}

Every closed path on this quiver (\ie\ every single trace gauge
singlet) may be deformed using the $F$-term constraints so that it
touches the $C$ node labelled $0$. We will consider the set of paths
on the quiver diagram based (\ie\ starting and ending) at the node
$0$. We will show that all such operators and therefore all single
trace operators in the chiral ring are generated by products of the
operators shown in Figure~\ref{Y32loops}.

First one may use the equations of motion to show that the
independent loops of Figure~\ref{Y32loops} are in 1-to-1
correspondence with the invariants in the GLSM. There are four
triangular paths $Y_{01}V_{12}^{1,2}U_{20}^{1,2}$, but the
${\partial W\over \partial Y_{10}}=0$ constraint sets
$V_{12}^{1}U_{20}^{1}=V_{12}^{2}U_{20}^{2}$ so there are three
independent triangles. These three paths correspond to the
invariants $abcd$, $abc^2$ and $abd^2$ in the GLSM. The two
``hourglass''-like operators $Y_{01}Y_{13}Y_{23}U_{20}^{1,2}$ correspond
to the invariants $a^3c$ and $a^3d$ in the GLSM. Finally, since the
lower indices are all fixed, denote the ``rabbit''-like operators
$Z_{05}U_{53}^{1,2}V_{34}^{1,2}U_{41}^{1,2}V_{12}^{1,2}U_{20}^{1,2}$
using only the upper indices as a sequence of five $1$s or $2$s as
in $Z11111, Z12111$ etc.
\begin{figure}
\bc \epsfig{file=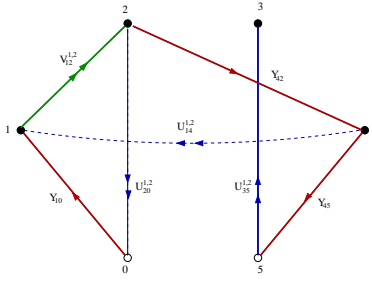, width=5cm} \caption{Decomposing loops
of the $Y^{32}$ quiver beginning with $Y_{01}V_{12}\ldots$ into its
irreducible loops.} \label{Y32gen1} \ec
\end{figure}
Using the equations of motion ${\p W\over\p Y}=0$, we may remove any
adjacent $2$s in an operator for example, $11221=11111$. We can also
shift $2$s by 2 places using the equations of motion, so
$12111=12221=11121$. Thus there are only six inequivalent
independent operators of this form\bea && Z21111=Z11211=Z11112\ ,
\qquad\quad Z11111\ , \qquad\quad
Z12111=Z11121\ , \nonumber\\
&& Z21211=Z21112=Z11212\ , \qquad\quad Z12121\ , \qquad\quad Z21212\
, \eea and they correspond to the six invariants $b^3c^{5-k}d^k,\
k=0,\ldots,5$ in the GLSM. So the full list of adjoint operators we
are considering is: $Y11, Y12, Y21, YYY1, YYY2$, and 6 operators of
the form $Z11111, Z12111$, etc. This set is in 1-to-1 correspondence
with the GLSM invariants.

\textbf{Generators}

Now we will show that all adjoints at node $0$ are generated by
products of these 11 generating loops. The superscripts will not be
important for this argument, so we will ignore them. Any operator,
$A$, adjoint at node $0$ must begin with either $Y_{01}$ or $Z_{05}$
since these are the only two outgoing links at node $0$. These links
must be similarly followed by similar outgoing links at either node
$1$ or node $5$ respectively, and so on through all nodes in the
loop, until the loop closes and returns to node $0$. We will try to
construct an operator that cannot be reduced to generators.
\begin{figure}
\bc \epsfig{file=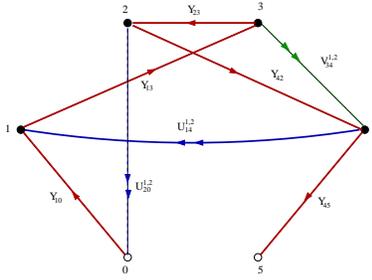, width=5cm} \caption{Decomposing loops
of the $Y^{32}$ quiver beginning with $Y_{01}Y_{13}\ldots$ into its
irreducible loops.} \label{Y32gen2} \ec
\end{figure}

We first consider operators beginning with $Y_{01}$ and illustrate
our argument in Figure~\ref{Y32gen1}. Suppose $A$ begins with
$Y_{01}V_{12}Y_{24}\ldots$. If we next insert $U_{41}$ then we have
$Y_{01} V_{12}Y_{24}U_{41}=(Y_{01} V_{12}U_{20})Y_{01}$ using the
${\partial W\over\partial V_{12}}=0$ equation, which is a generator
times another loop. So to generate something that cannot be reduced,
we must insert $Y_{45}U_{53}$ next.  But this yields $Y_{01}
V_{12}Y_{24}Y_{45}U_{53}$ which can be reduced to $Y_{01}
V_{12}Y_{24}U_{41}Y_{13}$ using the ${\partial W\over
\partial V_{34}}=0$ equation. But we just saw that if we leave node
$4$ along $U_{41}$, then the operator can be reduced using the
constraints. Thus if $A$ begins with $Y_{01} V_{12}Y_{24}$, then it
can be reduced.

Next, we suppose $A$ begins with $Y_{01}Y_{13}\ldots$ and illustrate
our argument in Figure~\ref{Y32gen2}. If we leave node $3$ along
$V_{34}$ then we have $Y_{01}Y_{13}V_{34}=Y_{01}V_{12}Y_{24}$ by the
${\partial W\over\partial U_{41}}=0$ equations. This brings us back
to the consideration of the previous paragraph, so instead we insert
$Y_{32}$ and then $Y_{24}$ since $Y_{01}Y_{13}Y_{32}U_{20}$ is one
of the generators shown in Figure~\ref{Y32loops}. Then after
inserting $Y_{24}$, we have $Y_{01}Y_{13}Y_{32}Y_{24}$ and we have
arrived at node $4$. Now if we leave node $4$ along $U_{41}$, then
we can use the ${\partial W\over\partial U_{41}}=0$ equation to
reduce the path to generators:
\begin{equation}
Y_{01}Y_{13}Y_{32}Y_{24}U_{41}=(Y_{01}Y_{13}Y_{32}U_{20})Y_{01}
\end{equation}
If instead we leave node $4$ along $Y_{45}$ to be followed by
$U_{53}^{1,2}$, then we can use the ${\partial W\over\partial
V_{34}^{1,2}}=0$ equation to replace the $Y_{45}U_{53}^{1,2}$ by
$U_{41}^{1,2}Y_{13}$:
\begin{equation}
Y_{01}Y_{13}Y_{32}Y_{24}Y_{45}U_{53}=Y_{01}Y_{13}Y_{32}Y_{24}U_{41}Y_{13}
\end{equation}
and we have just seen that this can be reduced.

\begin{figure}
\bc \epsfig{file=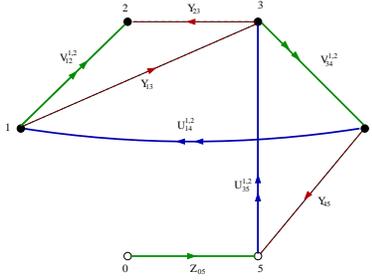, width=5cm} \caption{Decomposing loops
of the $Y^{32}$ quiver beginning with $Z_{05}U_{53}\ldots$ into its
irreducible loops.} \label{Y32gen3} \ec
\end{figure}

\textit{These arguments have shown that every loop beginning with
$Y_{01}$ can be reduced to the product of one of our 11 generators
and some other loop. Similar considerations, which are illustrated
in Figure~\ref{Y32gen3} show that every operator beginning with
$Z_{05}U_{53}\ldots$, may also be reduced to the product of one of
the generators and some other loop.}
Now iterating the results in the previous paragraph shows that all
paths adjoint at the C node $0$ can be reduced to a product of the
generators shown in Figure~\ref{Y32loops}.

\textbf{Commutation and GLSM relations}

We will now show that the 11 generators satisfy the same relations
as the invariants in the GLSM and that they commute. Consider first
the triangular generators shown in Figure~\ref{Y32loops}. The
correspondence with the GLSM is:
\begin{equation}
abc^2\sim Y21\ , \qquad abd^2\sim Y12\ , \qquad abcd\sim Y11=Y22\ .
\end{equation}
Then using the ${\partial W\over\partial V_{12}}=0$ equation and the
${\partial W\over\partial Y}=0$ equations, and denoting $Y_{24}$ by
$\bar{Y}$ and $U_{41}^{1,2}$ by $\bar{1}$ and $\bar{2}$, we have:
\begin{eqnarray}
Y21Y12&=&Y2\bar{Y}\bar{1}12=Y2\bar{Y}\bar{2}22=Y22Y22\ , \nonumber\\
Y12Y21&=&Y1\bar{Y}\bar{2}21=Y1\bar{Y}\bar{1}11=Y11Y11=Y22Y22\ ,
\end{eqnarray}
so that $[Y21,Y12]=0$ and $Y21Y12=(Y11)^2$ corresponding to
$abc^2abd^2=(abcd)^2$. Also:
\begin{eqnarray}
Y21Y11&=&Y2\bar{Y}\bar{1}11=Y2\bar{Y}\bar{2}21=Y22Y21=Y11Y21\ ,\nonumber\\
Y12Y22&=&Y1\bar{Y}\bar{2}22=Y1\bar{Y}\bar{1}12=Y11Y12=Y22Y12\ .
\end{eqnarray}
So we also have $[Y21,Y11]=[Y12,Y11]=0$, \ie\ the triangular
generators commute and satisfy their GLSM relations.

Now we consider commutation between the triangular and hourglass
shaped generators shown in Figure~\ref{Y32loops}. We start with
\begin{equation}
(Y_{01}Y_{13}Y_{32}U_{20}^a)(Y_{01}V_{12}^bU_{20}^c)\ .
\end{equation}
If $b=c$, then use the $F$-term constraints to set $a=c$. Now
proceed:
\begin{eqnarray}
(Y_{01}Y_{13}Y_{32}U_{20}^a)(Y_{01}V_{12}^bU_{20}^c)&=&Y_{01}Y_{13}
Y_{32}U_{20}^aZ_{05}U_{53}^{\bar{b}}Y_{32}U_{20}^c
\end{eqnarray}
where $\bar{b}$ is the opposite of $b$. Now use the ${\partial
W\over \partial Z_{05}}=0$ and the ${\partial W\over \partial
Y_{32}}=0$ equations to cyclically permute the label $a$ to the
final position. Then we can show that
\begin{eqnarray}
Y_{01}Y_{13}Y_{32}U_{20}^{\bar{b}}Z_{05}U_{53}^cY_{32}U_{20}^a
&=&(Y_{01}V_{12}^bU_{20}^c)(Y_{01}Y_{13}Y_{32}U_{20}^a)
\end{eqnarray}
so that $[Y_{01}Y_{13}Y_{32}U_{20}^a,Y_{01}V_{12}^bU_{20}^c]=0$. So the
hourglass and triangular generators commute.

Next we consider the commutation relations of the hourglass
generators and the rabbit generators involving $Z_{05}$:
\begin{eqnarray}
Z_{05}U_{53}^{1,2}V_{34}^{1,2}U_{41}^{1,2}V_{12}^{1,2}U_{20}^{1,2}
\end{eqnarray}
We will use the ${\partial W\over\partial U}=0$ and ${\partial
W\over\partial V}=0$ constraints here and since they preserve the
upper indices of the $U$s and $V$s (except the equations from
$U_{20}$ and $U_{53}$), we will allow the lower index structure to
be understood.

The $F$-term constraints imply the following relations for
$a,b,\ldots,g$ taking values 1 or 2:
\begin{eqnarray}\label{exch}
ZabcdeYfg=ZabYcdefg=ZaY\bar{b}Zcdefg=Y\bar{a}\bar{b}Zcdefg\ ,
\end{eqnarray}
If $fg=21$ and $ab=12$, then we obtain the commutation relations for
the operators corresponding to $[b^3d^{5-k}c^k,abc^2]=0$ for
$k=2,\ldots,5$ in the GLSM. \textit{eg}:
\begin{eqnarray}
(Z12111)(Y21)=(Y21)(Z11121)\ .
\end{eqnarray}
Similarly if we take $fg=12$ and $ab=21$, then we obtain the
relations $[b^3d^{5-k}c^k,abd^2]=0$ for $k=0,\ldots,3$ in the GLSM.
Using these relations and (\ref{exch}) we can show that the
relations corresponding to $[b^3d^4c,abc^2]=0$ and
$[b^3d^5,abc^2]=0$ follow. The commutation relations corresponding
to $[b^3d^{5-k}c^k,abd^2]=0$ follow in the same manner by using
(\ref{exch}) as above.

Setting $fg=11$ and $ab=11$ gives the relations corresponding to
$[b^3d^{5-k}c^k,abcd]=0$ for $k=1,\ldots,4$ in the GLSM and the two
remaining commutation relations here follow as before by using
commutation relations we have already proved:
\begin{eqnarray}
(Z12121)(Y11)&=&(Y21)(Z12111)=(Z12111)(Y21)=(Y11)(Z12121)
\end{eqnarray}
We have thus shown that all rabbit-like generators commute with all
triangular generators.

Certain relations from the GLSM can be recovered in gauge theory by
using (\ref{exch}). If we let $fg=12$, and $ab=12$, then we obtain
the four relations corresponding in the GLSM to
$(b^3d^{5-k}c^k)(abd^2)=(abc^2)(b^3d^{7-k}c^{k-2})$. Letting instead
$fg=11$ and $ab=12,21$, we obtain 10 more relations corresponding to
$(b^3d^{5-k}c^k)(abd^2)=(b^3d^{6-k}c^{k-1})(abcd)$ and
$(b^3d^{5-k}c^k)(abc^2)=(b^3d^{4-k}c^{k+1})(abcd)$ in the GLSM.

Finally we consider the commutation relations of the hourglass
operators and the rabbit like operators.  Letting indices of the $Y$
fields be understood and denoting the hourglass operators as $Y^31$
and $Y^32$, we can use the $F$-term constraints to show that
\be
(Y^3f)(Zabcde)=(Zf\bar{a}\bar{b}\bar{c}\bar{d})(Y^3e)\ .
\ee
Setting $a=e$ gives 10 of the hourglass-rabbit commutation relations
and the remaining two commutation relations follow from these and
previous relations. So all hourglasses commute with all rabbits.

The relations, corresponding to the 9 GLSM relations
$(b^3d^{5-k}c^k)(a^3c)=(abc^2)^{l_1}(abd^2)^{l_2}(abcd)^{l_3}$ where
$l_1+l_2+l_3=3$ and to the 9 relations
$(b^3d^{5-k}c^k)(a^3d)=(abc^2)^{l_1}(abd^2)^{l_2}(abcd)^{l_3}$ where
$l_1+l_2+l_3=3$ also now follow. So the generators in the gauge
theory reproduce the the GLSM invariants and all their relations
exactly.

\section{Singularities in $L^{abc}$}\label{sec:sing}

In this section, we will describe aspects of $L^{abc}$ theories
that have singular base 5-geometries. This happens if either of $a,b$
has common factors with either of $c,d$, as mentioned briefly earlier.

To see these singularities in the geometry ((see also \cite{knconiflips}),
we represent the full cone as a quotient (\ref{dterm}) as
$\left\{a|z_a|^2+b|z_b|^2=c|z_c|^2+d|z_d|^2\right\} //U(1)$. \
The orbit of the $U(1)$ gauge symmetry, with action\ $(z_a,z_b,z_c,z_d)
\to (e^{ia\theta}z_a,e^{ib\theta}z_b,e^{-ic\theta}z_c,e^{-id\theta}z_d)$,
is an $S^1$ which is covered once for $\theta\in(0,2\pi)$ in the region
where all $z_i$ are nonzero. But on surfaces where some of the
$z_i=0$, the $S^1$ may be wrapped more than once. For example,
suppose that $gcd(a,c)=h$. The $U(1)$ action on the surface
$z_b=z_d=0$, given by\ $a|z_a|^2=c|z_c|^2$, is $(z_a,0,z_c,0)\to
(e^{ik_1 h \theta}z_a,0,e^{-i k_2 h \theta}z_c,0)$ where $a=k_1h$ and
$c=k_2h$.  Then as $\theta$ goes from $0$ to $2\pi$, we wrap around
the $S^1$ $h$ times. So we have a $\BZ_h$ orbifold singularity
extending in the whole $z_b=z_d=0$ plane.

\begin{figure}
\bc
\epsfig{file=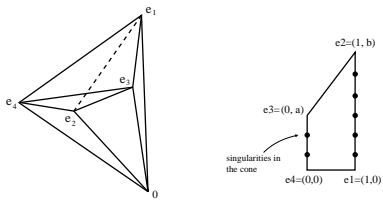, width=5cm}
\caption{The 3-dimensional toric fan for the singularity $Q_i$, defined
by lattice points $e_i$. Also shown is the plane containing the $e_i$
for the singular case $L^{a,b,a}$.}
\label{cone}
\ec
\end{figure}
The fact that we are dealing with toric singularities is useful, in
particular in recovering the above conditions on the singularities
of the base. The singularity (\ref{labcQ}) is represented by a toric
cone in an integral $\BN$ lattice defined by four lattice points
$e_i$ satisfying $\sum_iQ_ie_i=0$ (see Figure~\ref{cone}). The $e_i$
are coplanar (see \eg\ \cite{knconiflips}) iff $\sum_i Q_i = 0$,
\ie\ $a+b=c+d$, in which case these are supersymmetric cones. Then
the $e_i$ may be written in the form $e_i=(1,w_i)$ so that we can
draw the four vectors on a 2-plane. This is not unique since
applying an $SL(3,\BZ)$ transformation to the vectors $e_i$ yields a
cone describing the same geometry: a representation for the $e_i$ is
(suppressing the third coordinate) \be\label{coneei} e_4=(0,0) ,
\qquad e_1=(1,0) , \qquad e_2=(-al,c) \ , \qquad e_3=(ak,b)\ ,
\ee
where $k,l$ are two integers satisfying $bl+ck=1$ (we assume $b,c$
are coprime). The total $\BN$ lattice volume of the cone is
$V_{cone}=a+b\ (=c+d)$, giving the number of gauge groups $N_g$ in
the quiver. A toric cone of volume $V_{cone}>1$ is
singular\footnote{See \eg\ \cite{knconiflips} for discussions on the
relevance of these $\BN$ lattice volumes and resolution of
singularities in the context of nonsupersymmetric conifold-like
singularities with closed string tachyons.}, so that there are $a+b$
subcones in the interior of the cone: subdividing by lattice points
either in the interior of the cone or on the ``walls'' (faces) gives
$a+b$ subcones each of volume $V_{subcone}=1$, \ie\ a complete
resolution representing a smooth space.

The fan encodes information on when the singularity is isolated
(pointlike). The singularity (\ref{labcQ}) is isolated, \ie\ the base
space $X$ is smooth, if the toric fan does not contain any lattice
points on its walls \cite{lerman} (see also \cite{knconiflips} in
the context of these conifold-like singularities). This is algebraically
equivalent to
the condition that each of $a,b$ is coprime with each of $c,d$. It is
easy to illustrate this if one of the $a,b,c,d$, say $b=1$. Then from
the fan, we see that there are no lattice points on the $\{e_2,e_4\}$
wall if $a,c$ are coprime: if there is a common factor $h$, then we
can construct the $h-1$ lattice points\ \
${1\over h}\ [ke_2+(h-k)e_4] , \ k=1,\ldots,h-1$,\
which lie on the $\{e_2,e_4\}$ wall. These correspond to the
$h-1$ twisted sector states of the $Z_h$ orbifold singularity described
earlier. Similarly, there are no lattice points on the $\{e_2,e_3\}$
wall if $a,1-c$ are coprime, \ie\ $a$ is coprime with $d=a+1-c$.
Doing this more generally shows that these are exactly the conditions
we saw earlier for the orbifold singularities in the geometry.

\subsection{$L^{aba}$}

\begin{figure}
\bc
\epsfig{file=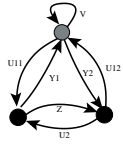, width=1.5cm}
\caption{The quiver for $L^{121}$}
\label{L121}
\ec
\end{figure}
This family of geometries is given by
$Q = ( \bA{cccc} a & b & -a & -b \eA )$,
with $a,b$ coprime: in this case, a basis of invariant monomials is\
$x_1=z_az_c, \ x_2=z_a^bz_d^a, \ x_3=z_b^az_c^b, \ x_4=z_bz_d,$
satisfying the relation\ $x_1^bx_4^a=x_2x_3$.
The toric cone for the geometries $L^{aba}$ is defined by the vectors
$e_i=\{(0,0)\ ,(1,0)\ , (0,a)\ ,(1,b)\}$ (see Figure~\ref{cone}).
The lattice points on the faces of the cone reflect the fact that
these spaces are singular on the base.

\begin{figure}
\bc
\epsfig{file=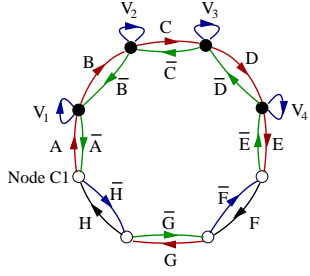, width=4cm}
\caption{The quiver for $L^{aba}$}
\label{Laba}
\ec
\end{figure}

We consider here $L^{aba}$ with $gcd(a,b)=1$. This is a singular
geometry with two orbifold singularities of order $a$ and $b$. The
quiver diagram of the gauge theory is circular with $2a$ type $C$
nodes and $(b-a)$ type $A$ nodes (see Figures~\ref{L121}, ~\ref{Laba}).
For our purposes, the charges of the fields will not be important, so
we will label the fields on the quiver as
$\{\bar{A},A,\bar{B},B,\bar{C},C,\ldots\}$ and $V_i$ on node $i$ if
is a type $A$ node.  We consider $L^{232}$ as an example. The
general case is an obvious extension. We chose four adjoint
generators at $C_1$:
\begin{equation}
x=A\bar{A} \qquad y=\bar{E}E \qquad s=ABCDE \qquad
t=\bar{E}\bar{D}\bar{C}\bar{B}\bar{A}
\end{equation}
The commutation relations follow as in the non-singular cases:
\begin{equation}
xs=A\bar{A}ABCDE=ABVCDE=ABCD\bar{D}DE=ABCDEA\bar{A}=sx
\end{equation}
These four commuting adjoints at $C_1$ generate all gauge invariants
passing through $C_1$ satisfying the $F$-term constraints. We can
also see that these four adjoints satisfy $x^2y^3=st$:
\begin{eqnarray}
x^2y^3&=&(A\bar{A})^2(\bar{E}E)^3=A\bar{A}A
B(\bar{B}B)^2\bar{B}\bar{A}=ABV(\bar{B}B)^2\bar{B}\bar{A}=
ABVC(\bar{C}C)\bar{C}\bar{B}\bar{A}\\
&=&ABCD\bar{D}\bar{C}C\bar{B}\bar{A}=ABCDE\bar{E}\bar{D}\bar{C}
\bar{B}\bar{A}=st
\end{eqnarray}
These four generators reproduce the geometry completely, up to
complex structure, but they do not generate all gauge invariant
states of the gauge theory. Any path that touches three nodes and at
least two neighboring $C$ nodes, can be moved around the graph using
the $F$-term constraints, until it touches $C_1$. Therefore, all
such paths are generated by the four adjoints $x,y,s,t$.  Any path
touching $A$ nodes, but not touching two neighboring $C$ nodes can
be moved to the loop wrapping only $A\bar{A}$ by using the
constraints. This path touches $C_1$ and is also generated by the
adjoints $x,y,s,t$. Paths that touch only 2 $C$ nodes, however,
cannot be deformed at all using the constraints, so we must count
these operators separately.

\textit{The single trace gauge invariants are counted by traces of
words made from $x,y,s,t$ satisfying $x^2y^3=st$ and the additional
singlets $w_{n_1}=(C\bar{C})^{n_1}$, $u_{n_2}=(D\bar{D})^{n_2}$,
$v_{n_3}=(V)^{n_3}$.}

In the general $L^{aba}$ theory, with $gcd(a,b)=1$, paths touching
$C_1$ are all generated by the four adjoints $x,y,s,t$ satisfying a
constraint $x^by^a=st$, where $x$ is the loop connecting $C_1$ to
$A_1$, $y$ is the loop between $C_1$ and $C_2$ and $s,t$ are the
loops around the whole graph in each direction. To demonstrate this
relation, we label the lines before the $i^{th}$ gauge group as
$A_i,\bar{A_i}$, and label the lines between $C$ nodes as
$C_0,\ldots, C_{2(a-1)}$. Then we start with a set of loops at $C_1$
and we move them around the circle in the clockwise direction.  We
have: {\small
\begin{eqnarray}
(A_1\bar{A}_1)^{b}(\bar{C}_0C_0)^{a}&&=A_1(\bar{A}_1A_1)^{b-1}
V_1^{a}\bar{A}_1
=A_1A_2(\bar{A}_2A_2)^{b-2}\bar{A}_2V_1^{a}\bar{A}_1\nonumber\\
&&=A_1A_2(\bar{A}_2A_2)^{b-2}V_2^{a}\bar{A}_2\bar{A}_1\nonumber\\
&&=A_1A_2\ldots A_{b-a}A_{b-a+1}(\bar{A}_{b-a+1}A_{b-a+1})^{a-1}
\bar{A}_{b-a+1}V_{b-a}^{a}\bar{A}_{b-a}\ldots
\bar{A}_2\bar{A}_1\nonumber\\
&&=A_1A_2\ldots A_{b-a}A_{b-a+1}(C_{2(a-1)}
\bar{C}_{2(a-1)})^{2a-1}\bar{A}_{b-a+1}\bar{A}_{b-a}\ldots \bar{A}_2
\bar{A}_1\nonumber\\
&&=A_1A_2\ldots A_{b-a}A_{b-a+1}C_{2(a-1)}C_{2(a-1)-1}\ldots
C_0\nonumber\\
&&\qquad\cdot\ \bar{C}_0\ldots \bar{C}_{2(a-1)-1}
\bar{C}_{2(a-1)}\bar{A}_{b-a+1}\bar{A}_{b-a}\ldots
\bar{A}_2\bar{A}_1 \nonumber
\end{eqnarray}
}
The generators $x,y,s,t$ do not generate all gauge invariant
operators. We must also count the singlets
$\omega_n^i=(C_i\bar{C_i})^n$ where $i=1,2,\ldots 2(a-1)$
corresponding to $C\bar{C}$ loops. Finally, given any path involving
$V_i$, we can always remove $V_i$ from the path by using the
$F$-term constraint equations if $V_i$ is contracted with any other
line. So the only words including $V_i$ which we must count are the
singlets $v_k^i=\tr V_i^k$, $i=1,\ldots b-a$.

There are $N_g-2$ extra singlet operators which are not generated by
$x,y,s,t$. These are\ $\omega_n^i=(C_i\bar{C_i})^n$, where
$i=1,2,\ldots, 2(a-1)$ and $v_k^i=\tr V_i^k$, $i=1,\ldots, b-a$. The
toric fan shows $a-1+b-1=N_g-2$ lattice points on the faces,
corresponding to $N_g-2$ orbifold-like singularities, as we have
described earlier. Thus we would expect that these extra operators
in the gauge theory may be interpreted in the bulk as $N_g-2$
twisted sector closed string states. $L^{aba}$ has two orbifold-like
singularities of order $a$ and $b$ respectively. These singularities
have complex dimension one and twisted sector states in the bulk are
restricted to move on these singularity planes. In other words, we
expect such states to rotate only in this plane (\ie\ angular
momentum only normal to the plane): thus the charges of the twisted
sector states associated to the $z_i$ normal to the singularity
plane should be zero. Further, the charges associated to the $z_i$
covering the singularity plane should be given by the order of the
orbifold singularity. In the gauge theory we have $a-1$ $C\bar{C}$
type operators of the form $YU_1$ and $a-1$ of the form $ZU_2$. This
reproduces the spectrum of twisted states, with charges given in
Table~\ref{singcharge}. {\small
\begin{table}[htb]
\bc
\caption{$L^{aba}$: charges of operators}
\label{singcharge}
\begin{tabular}{|c|c|c|}
\hline
Field & Charge & Multiplicity \\ \hline\hline
$V_i$ & $(0,b,0,-b)$ & $b-a$  \\ \hline
$ZU_2$ & $(0,b,0,-b)$ & $a-1$ \\ \hline
$YU_1$ & $(a,0,-a,0)$ & $a-1$ \\ \hline
\end{tabular}
\ec
\end{table}
}

The partition function over ${1\over 2}$-BPS states will include
these localized or twisted sector closed string states propagating
in $AdS_5\times X^5$. It is clear from the discussion here that the
partition function of the chiral ring will factor into the form:
\begin{equation}
Z_{bps}=Z_{gg}Z_{twist}
\end{equation}
where $Z_{gg}$ is the partition function appearing in
(\ref{multipf}) and $Z_{twist}$ is the partition function over the
twisted states. Since $\tr V_i^{N+1}$ is related to traces of lower
powers by trace relations, we have exactly $N$ independent harmonic
oscillators, $\tr V_i^k,\ k=1\ldots N$, for each $i$, giving
\begin{equation}
Z_{twist}=\prod_{i=1}^{b-a}\prod_{n=1}^{N}{1\over 1-v_i^n}
\end{equation}
where $v_i$ carries the charges of $V_i$.

It is interesting to wonder whether the whole chiral ring partition
function might be recovered from a single bulk giant graviton
quantization. In this context it is natural to imagine that twisted
states with energies of order $N$ or higher puff up into D3-branes,
analogous to giant gravitons, suggesting a generalization of
\cite{mikhailov} to include these localized giant gravitons.

\section{Discussion}

We have obtained the mesonic chiral ring partition function in \No\
theories arising on the worldvolumes of D3-branes stacked at $Y^{pq},
L^{abc}$ Calabi-Yau conical singularities. This agrees with their bulk
interpretation as cohomologically trivial giant gravitons, suggesting
a non-renormalization theorem for ${1\over 2}$-BPS states in these
theories. The quiver theories for $L^{aba}$, with singular base
geometries, contain extra operators not counted by the naive bulk
partition function: these have a natural interpretation in terms of
twisted states localized at the orbifold-like singularities in the
bulk. It would be interesting to understand possible giant graviton
interpretations of these localized states, possibly generalizing
\cite{mikhailov}.

Our field theory techniques should apply to other quiver
theories, such as those arising from D3-branes at $\BC^3/\Gamma$
orbifold singularities (which are also toric), and perhaps more
generally. The preliminary results of \cite{wip} suggest that our
approach generalizes to baryonic operators in the quiver,
corresponding to D3-brane giant gravitons wrapped on the
non-contractible 3-cycle in the bulk.

It is worth mentioning that the counting problem addressed here shows
agreement between the weak and strong coupling calculations even for
large numbers of D3-brane giant gravitons. This suggests an underlying
topological structure to the \No\ chiral ring, preserved under the
backreaction of the giant gravitons. For large numbers of giant
gravitons, it would of course seem natural to cross over to a
supergravity description, generalizing LLM \cite{llm} to geometries
with $L^{abc}$ asymptotics: see \cite{narain} for some recent progress
in this context. Quantization of the LLM solutions reproduced the
large $N$ spectrum of $1/2$ BPS operators in $\mathcal{N}=4$ SYM
\cite{Grant},\cite{Maoz} and a similar quantization of LLM type
solutions in $L^{abc}$ spaces might be expected to reproduce the large
$N$ limit of the chiral ring partition functions in this paper. It
would also be interesting to generalize the approach of
\cite{berensteinE} to shed light on matrix models and the emergence of
geometry in the context of the \No\ theories considered here.

\vspace{10mm}
{\small {\bf Acknowledgments:} It is a great pleasure to thank Shiraz
Minwalla for innumerable discussions on this paper and collaboration
on \cite{wip}, as well as Davide Gaiotto for various discussions and
collaboration in \cite{wip}. We would also like to thank Subhaneil
Lahiri and Gautam Mandal for discussions.}

\end{document}